\documentclass[10pt,letterpaper]{article}
\usepackage[top=0.85in,left=1.5in,footskip=0.75in,marginparwidth=2in]{geometry}
\usepackage{amsmath}

\usepackage[utf8]{inputenc} 
\usepackage[T1]{fontenc} 
\usepackage{lmodern} 
\usepackage{graphicx}
\usepackage{subcaption}
\usepackage{mathtools}
\usepackage{verbatim}
\usepackage{amssymb}
\usepackage{gensymb}

\usepackage{cite}

\usepackage{nameref,hyperref}

\usepackage{microtype}
\DisableLigatures[f]{encoding = *, family = * }

\usepackage{changepage}

\usepackage[aboveskip=1pt,labelfont=bf,labelsep=period,singlelinecheck=off]{caption}

\usepackage{lastpage,fancyhdr,graphicx}
\usepackage{epstopdf}
\fancyheadoffset[L]{2.25in}
\fancyfootoffset[L]{2.25in}

\usepackage{color}

\definecolor{Gray}{gray}{.25}

\usepackage{graphicx}

\begin{document}
\vspace*{0.35in}

\begin{flushleft}
{\LARGE
\textbf\newline{Direct Frequency-Mode-Stable Laser Amplification at Terahertz Burst Rates}
}
\newline

\textbf{Vinzenz Stummer\textsuperscript{1,*},
Tobias Flöry\textsuperscript{1},
Matthias Schneller\textsuperscript{1},
Edgar Kaksis\textsuperscript{1},
Markus Zeiler\textsuperscript{1},
Audrius Pugžlys\textsuperscript{1,2},
Andrius Baltuška\textsuperscript{1,2}
}
\\
\bigskip
[1] Photonics Institute, TU Wien, Gusshausstrasse 27/387, 1040 Vienna, Austria
\newline
[2] Center for Physical Sciences \& Technology, Savanoriu Ave. 231 LT-02300 Vilnius, Lithuania
\\
\bigskip
* vinzenz.stummer@tuwien.ac.at

\end{flushleft}

\section*{Abstract}
Generation of high-fidelity amplified pulse bursts with a regular interpulse interval yields, in the spectral domain, an equidistant pattern of narrowband spectral modes, similar to frequency combs produced by cw mode-locked lasers, but with greatly increased pulse energy. Despite their great potential for nonlinear spectroscopy, material processing, etc., such long frequency-stable bursts are difficult to generate and amplify because of prominent temporal intensity modulation even after strong dispersive pulse stretching. This study presents a burst generation method based on a master-oscillator regenerative-amplifier system that allows for chirped-pulse amplification (CPA) with high scalability in pulse number. A gradual smoothing of temporal intensity profiles at an increasing number of pulses is discovered, demonstrating an unexpected recovery of the CPA performance at terahertz (THz) intraburst repetition rates. In consequence, a self-referenced stable burst spectral peak structure with megahertz (MHz) peak width is generated, without risk of amplifier damage caused by interference of chirped pulses. This result eliminates limitations in burst amplification and paves the way for advancements in ultrashort-pulse burst technology, particularly for its use in nonlinear optical applications.

\section{Introduction}
\begin{figure}
\includegraphics[width=1.0\linewidth]{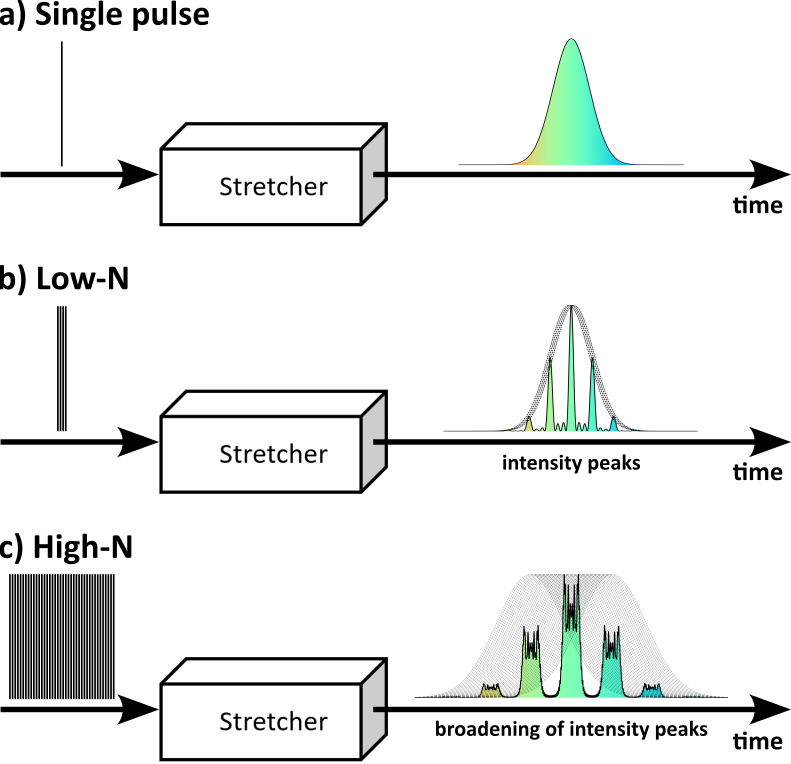}
\caption{\textbf{Depiction of the effect of temporal intensity peak broadening for chirped THz-rate bursts at increasing pulse numbers.} After the stretcher, the pulse spacing is much smaller than the chirped pulse duration. Temporal scaling before and after the stretcher differ for visualization purposes. \textbf{a) Single-pulse:} High-frequency, blue components advance lower-frequency, red components in time. For a linear chirp the spectrum is translated into the time domain. \textbf{b) Low-$N$ regime:} When chirping only a handful of pulses closely together, burst-typical spectral peaks are mapped into the time domain. \textbf{c) High-$N$ regime:} As soon as the pulse number becomes large enough, there is no clear overlap in time of all the pulses anymore. The spectral peaks are not mapped into the time domain and temporal intensity peaks are broadened.\label{fig:intro}}
\end{figure}

Finite trains of ultrashort pulses, also commonly known as ultrashort-pulse bursts, are becoming increasingly relevant in controlling the orientation and alignment of molecules \cite{salomon_optimal_2005,loriot_laser-induced_2007,koch_quantum_2019,junnemann_alignment_nodate}, the generation of plasma waves \cite{umstadter_nonlinear_1994}, electron bunch generation \cite{boscolo_generation_2007} and amplification \cite{sudar_burst_2020}, or material ablation \cite{kerse_ablation-cooled_2016}. This development is originating from fundamental differences in the response of a system when excited with a burst-mode format in contrast to a single pulse. Such difference is also known when comparing the use of a single ultrashort pulse with that of an optical comb in spectroscopic applications \cite{picque_frequency_2019}. In these, the spectral response may span the entire bandwidth of the comb with very narrow, down to sub-millihertz \cite{schibli_optical_2008}, linewidths achieved with an infinite train of pulses. An ultrashort-pulse burst may consist of $N$ pulses with a given burst rate $1/\Delta t$, where $\Delta t$ is the interpulse spacing. The linewidth is given by the inverse duration of the burst $1/(N\Delta t)$ with a linewidth spacing equal to the burst rate $1/\Delta t$. The obtainment of spectral peaks with largest intensities can thus be achieved by generation of a burst with the highest number of ultrashort pulses (high-$N$) at terahertz (THz) burst rates, with the interpulse spacing comparable to the ultrashort pulse duration. By this, a spectrum with only few narrow lines, which contain the whole burst energy, is obtained. While bursts with GHz-rates, or lower, are an established technology \cite{park_mechanisms_2023, kalaycioglu_high-repetition-rate_2018, sudar_burst_2020}, the concentration of energy on a small number of spectral lines at THz burst rates opens up opportunities for nonlinear optical applications such as Stimulated Raman Scattering (SRS) \cite{prince_stimulated_2017} or Resonantly-Enhanced Multi-Photon Ionization (REMPI) \cite{shneider_laser_2012} that demand both, high peak power but also spectral selectivity. While the first prerequisite cannot be met by frequency combs due to their low peak power compared to a single pulse, the latter criterion cannot be fulfilled when being limited in pulse number in burst-mode generation. However, existing burst-mode systems have exactly this problem of pulse number scalability when generating ultrashort-pulse bursts at THz intraburst repetition rates. In this regime, common burst generation techniques rely on single-pulse division by using pulse shapers \cite{efimov_adaptive_1998,liu_nonlinear_1995,pastirk_no_2006,ahn_terahertz_2003,dugan_high-resolution_1997,hacker_micromirror_2003}, beam splitters \cite{muller_scaling_2021}, birefringent crystals \cite{radzewicz_passive_1996,zhou_efficient_2007,dromey_generation_2007}, or nested Mach-Zehnder interferometers \cite{bitter_generating_2016}. Independent of the approach, the practical limit in pulse number is ten to maximally a few hundred of pulses, due to the $1/N$ energy throughput of pulse division techniques \cite{bitter_generating_2016}. With the burst comprising pulses that are ultrashort, amplification techniques rely on Chirped-Pulse Amplification (CPA) \cite{strickland_compression_1985} that raises several problems in this regime. The most crucial one is the appearance of interference effects due to pulses that are much more strongly chirped than their individual spacing. By this, the temporal intensity profile of a stretched burst resembles the burst spectrum with a narrow peak structure that is dictated by the burst rate (See Fig. \ref{fig:intro}b). Another problem is caused by instabilities of the pulse spacing, or equivalently, the pulse-to-pulse phase slip $\phi_s$, leading to drifts of the spectral peaks. Given all these complications, amplification of ultrashort-pulse THz-rate bursts relied so far on individual pulse-phase modulation in order to suppress these interference effects in the time-domain, and on phase-slip stabilization techniques to allow for a stable peak structure \cite{stummer_programmable_2020}. \\
In this work, we develop a numerical model and analyze its results, and validate experimentally the time-frequency properties of ultrashort-pulse THz-rate bursts at high ($N \gg 10$) pulse numbers. We demonstrate a system that builds on a master-oscillator regenerative-amplifier setup with only minor modifications to its single-pulse operation to allow for burst-mode operation. By using direct time-domain methods for burst generation based on in-loop accumulation of oscillator pulses, we are not limited in seed burst energy. Our proposed method is therefor easily scalable in pulse number, up to ten thousands of pulses, in contrast to existing methods based on pulse division. Further, it allows to generate a stable burst spectral peak structure by direct stabilization, without any need of an external reference. An operating regime corresponding to an intrinsic smoothing of the temporal intensity profile of chirped THz-rate bursts is identified at high pulse numbers (See Fig. \ref{fig:intro}c). This burst amplification regime gives a sustainable energy extraction from an amplifier, even at high ($N\gg10$) pulse numbers without risk of intensity-induced optical damage. Further than that, for burst durations larger than the chirped pulse duration, the extractable energy of an amplifier can be shown to be increased. For such high pulse numbers the proposed technique combines CPA and Divided Pulse Amplification (DPA) \cite{zhou_divided-pulse_2007} in an unprecedented way. 

\section{Analytical Description}
\label{sec:theory}
We first describe all phenomena in an analytical way to outline the underlying physics. For the further interested reader, we provide derivations of the results communicated in this chapter in the Supplementary.
\subsection{Single pulse}
When chirping a pulse by a common prism or grating stretcher, a frequency-dependent phase is imposed on a pulse. In this work, we assume linearly chirped Gaussian pulses that can be described in the frequency domain as \cite{diels_ultrashort_2006}

\begin{align}
    \label{eq:theory-Ep_omega}
    \tilde{E}_P(\omega+&\omega_0) =\nonumber\\
        &\tilde{E}_{0}
        \exp{\left[
        -(1+iC)
        \left(
        \sqrt{2\ln{(2)}}\frac{\omega}{\omega_{FWHM}}
        \right)^2
        \right]
        }
\end{align}

with complex amplitude $\tilde{E}_{0}$, chirp parameter $C$, central frequency $\omega_0$ and Full-Width-at-Half-Maximum (FWHM) bandwidth $\omega_{FWHM}$.\\
The imposition of a chirp leads to a broadening of the pulse in time

\begin{equation}
    \label{eq:theory-tau}
    \tau_{FWHM} =
    \frac{
    4\ln{(2)}\sqrt{1+C^2}
    }
    {\omega_{FWHM}},
\end{equation}

with $\tau_{FWHM}$ being the FWHM pulse duration.
In this description, the field is given in the time-domain by

\begin{align}
    \label{eq:theory-Ep_t}
    E_P&(t) =\nonumber\\ 
    &E_{0}\exp{
    \left[
    -i\omega_0 t
    -(1+iC)\left(\sqrt{2\ln{(2)}}\frac{t}{\tau_{FWHM}}\right)^2
    \right]
    },
\end{align}

with a complex amplitude $E_{0}$. 

\subsection{Ultrashort-Pulse Bursts}
In the following, we define an ultrashort-pulse burst as a finite train of ultrashort pulses that are closely spaced in comparison to their duration. For describing ultrashort-pulse bursts analytically, we assume that all pulses are, up to a constant pulse-to-pulse phase slip $\phi_s$, equal. The analytical formulation of a burst consisting of $N$ such pulses with a $\Delta t$ interpulse spacing is then given in the time domain by a summation over the pulse fields $E_{n}(t)$ as following

\begin{align}
    E_B(t) &= \sum_{n=0}^{N-1} E_{n}(t) \label{eq:n-pulses}\\
    &= \sum_{n=0}^{N-1} E_{P}(t-n\Delta t) \exp{(in\phi_s)},
\end{align}

with $E_{P}(t)$ as in Eq. \ref{eq:theory-Ep_t}.\\
The corresponding description in the frequency domain is given by

\begin{align}
    \tilde{E}_B(&\omega+\omega_0) =\nonumber\\ 
    &\tilde{E}_P(\omega+\omega_0) \cdot \sum_{n=0}^{N-1}\exp{\left(-in(\Delta t (\omega+\omega_0) - \phi_{s})\right)}.
    \label{eq:Eb_omega}
\end{align}

The Wigner distribution gives useful insights, with a full time-frequency picture of a complex-valued signal. For the burst field, it is \cite{diels_ultrashort_2006}

\begin{equation}
    \mathcal{W}_B(t,\omega) \coloneqq \int_{-\infty}^{\infty} E_B(t+s/2)E_B^*(t-s/2)\exp{(-i\omega s)}ds.
    \label{eq:wigner-def}
\end{equation}

For the $N$-pulse field (Eq. \ref{eq:n-pulses}), the Wigner distribution consists of $N$ signal terms $\mathcal{W}_{B,n}^{(S)}(t,\omega)$ and $N(N-1)/2$ interpulse interference terms $\mathcal{W}_{B,nm}^{(I)}(t,\omega)$ \cite{hlawatsch_interference_1997}

\begin{equation}
    \mathcal{W}_B(t,\omega) = \sum_{n} \mathcal{W}_{B,n}^{(S)}(t,\omega) + \mathop{\sum_{n}\sum_{m}}_{m>n} \mathcal{W}_{B,nm}^{(I)}(t,\omega)
\end{equation}

Of further interest are the Wigner marginal integrals \cite{diels_ultrashort_2006}, i.e. the integration over the time axis which gives the spectrum $S(\omega)$ and the integration over the frequency axis, which gives the intensity in time $I(t)$:

\begin{subequations}
    \begin{equation}
        S(\omega) = \frac{1}{2\sqrt{(\mu_0/\epsilon)}}\int_{-\infty}^{\infty}
        \mathcal{W}_B(t,\omega)dt
    \end{equation}
    \begin{equation}
        I(t) = \frac{1}{2\sqrt{(\mu_0/\epsilon)}}\int_{-\infty}^{\infty}
        \mathcal{W}_B(t,\omega)d\omega
    \end{equation}
    \label{eq:wigner-marginals}
\end{subequations}

In Eq. \ref{eq:wigner-marginals}, $\mu_0$ is the vacuum permeability and $\epsilon$ is the electric susceptibility. We note at this point that the Wigner distribution, as given in Eq. \ref{eq:wigner-def}, is always a real-valued distribution with no imaginary part. 

\section{Numerical Results}
\label{sec:numerical}
We calculate numerically the Wigner distribution $\mathcal{W}_B(t,\omega)$ for an ultrashort-pulse burst, according to the definition given in Eq. \ref{eq:wigner-def}, with an intraburst pulse spacing of 1 ps. Depending on the FWHM pulse duration $\tau_{FWHM}$ (by variation of the chirp parameter $C$) and on the compressed burst duration $(N-1)\Delta t$ (by variation of the pulse number $N$), we are able to identify three regimes: compressed pulses (250 fs pulse duration), few strongly chirped pulses (200 ps, 10 pulses) and many strongly chirped pulses (200 ps, 80 pulses). Further, we show only calculations with zero phase slip $\phi_s=0$ in the main text. 

\begin{figure*}
\centering
    \begin{subfigure}[htbp]{\textwidth}
    \centering
    {\includegraphics[width=1.0\textwidth]{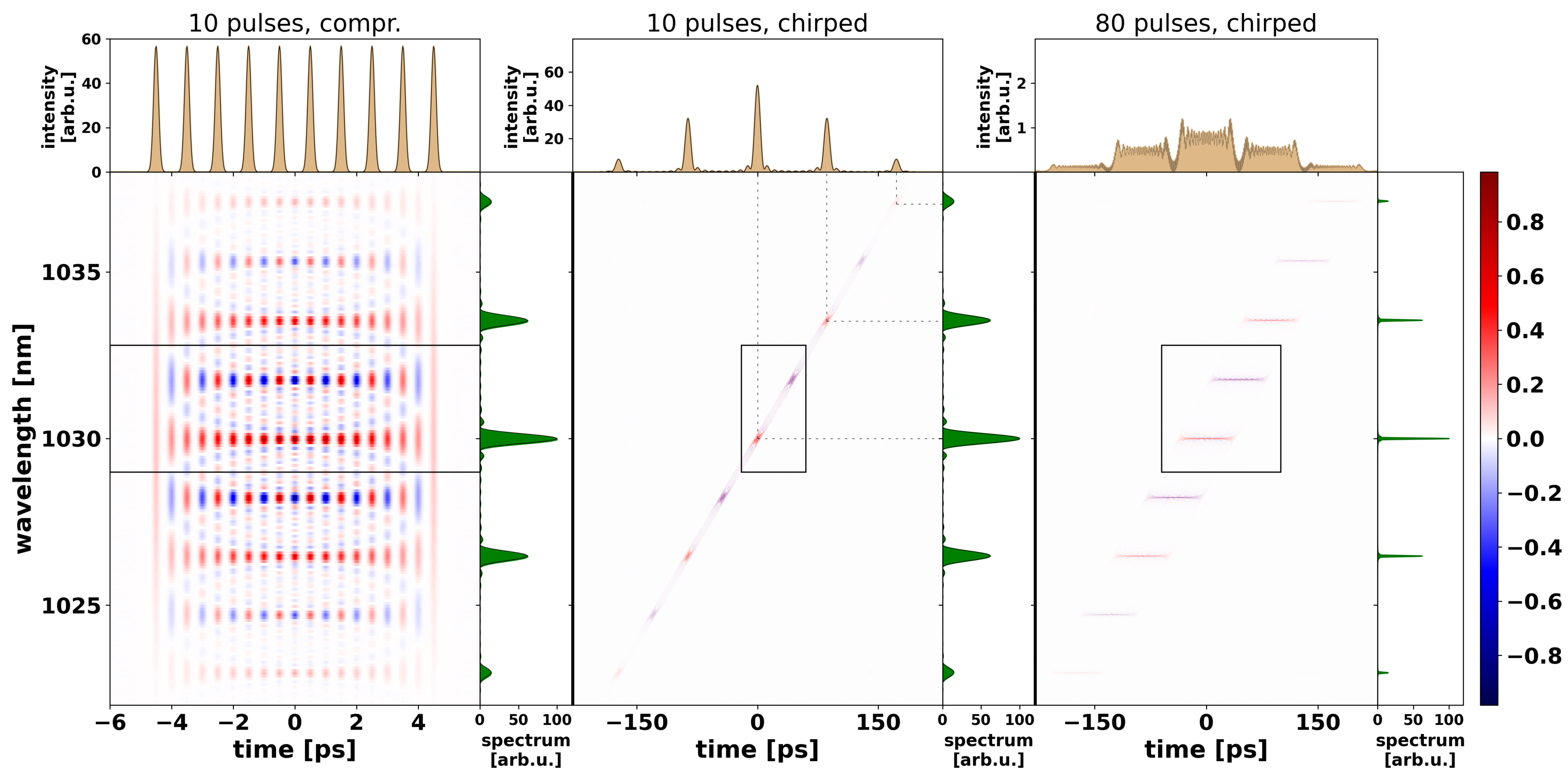}}
    \caption{Real part of the Wigner distribution $\mathcal{W}_B(t,\lambda)$ and its marginal distributions (intensity $I(t)$ and spectrum $S(\lambda)$). The mapping between spectral and temporal peaks is marked by dashed lines (middle plot).}
    \label{fig:wigner}
    \end{subfigure}

    \begin{subfigure}[htbp]{\textwidth}
    \centering
    {\includegraphics[width=1.0\textwidth]{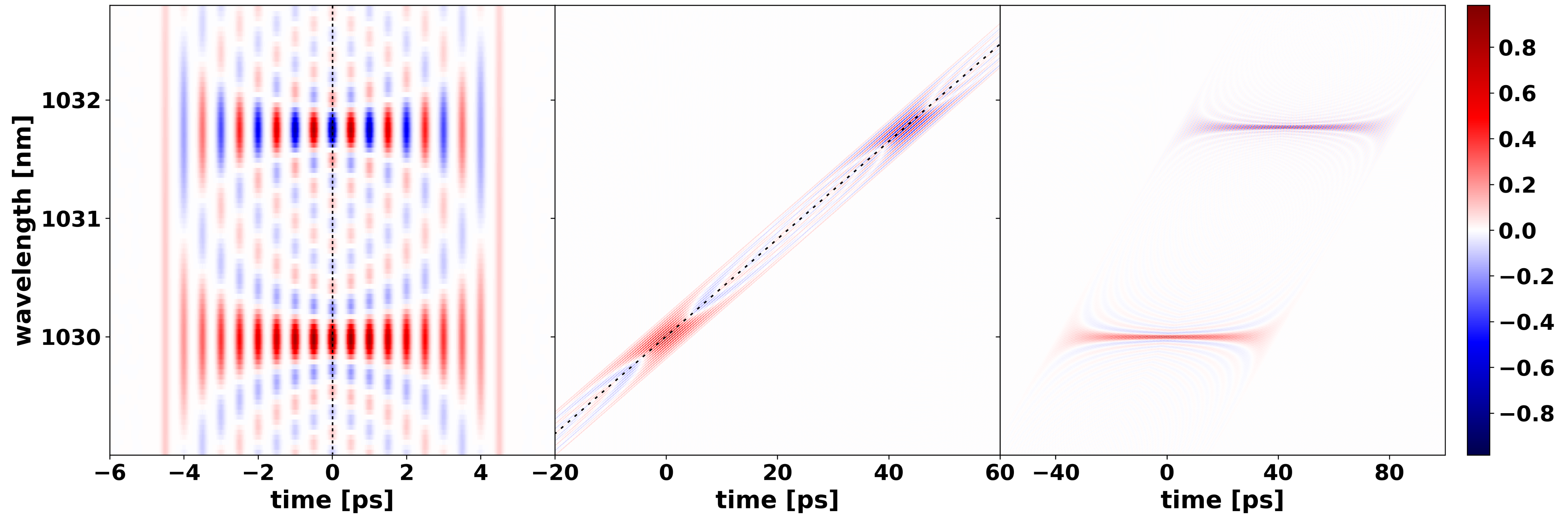}}
    \caption{Zoom into the Wigner distribution $\mathcal{W}_B(t,\lambda)$ as indicated by the black rectangles in (a). The symmetry lines of the 10-pulse Wigner distributions are marked by dashed lines.}
    \label{fig:wigner-zoom}
    \end{subfigure}
    
    \caption{\textbf{Numerical Wigner calculation results} consisting of the calculated real part of the Wigner distribution $\mathcal{W}_B(t,\lambda)$ (red and blue) and (a) its marginal distrbutions, i.e. intensity $I(t)$ (light brown) and spectrum $S(\lambda)$ (green), and (b) a zoom into the Wigner distribution for a better visualization of the periodic pattern. We represent the results depending on wavelength $\lambda$, instead of frequency $\omega$, for a better comparison with the experimental results shown later in this article. Calculations were performed with a burst with 8.5 nm bandwidth at 1030 nm (corresponding to a compressed pulse duration of 250 fs) with a pulse spacing of 2 ps. Left: 4 pulses compressed ($C=0, \tau_{FWHM}=250$ fs). Middle: 4 pulses chirped ($C=1.66\cdot10^7, \tau_{FWHM}=200$ ps). Right: 40 pulses chirped ($C=1.66\cdot10^7, \tau_{FWHM}=200$ ps).}
    \label{fig:wigner-all}
\end{figure*}

\subsection{Compressed Pulses}
\label{sec:numerical-compr}
The chirp parameter $C$ is negligible and the pulse duration $\tau_{FWHM}$ is smaller than the pulse spacing $\Delta t$ 

\begin{subequations}
\begin{equation}
    C\approx0,
\end{equation}
\begin{equation}
    \tau_{FWHM}<\Delta t.
\end{equation}
\end{subequations}

This is the typically known case prior, or after, Chirped-Pulse Amplification (CPA) where all pulses are compressed and well separated from each other in time (Fig. \ref{fig:wigner-all}, left side). For the first ($n=0$) and last ($n=N-1$) pulse, the Wigner distribution consists only of non-oscillatory positive-valued signal terms $\mathcal{W}_{B,n=0/n=N-1}^{(S)}(t,\omega)$. For each pair ($n,m$) of pulses, that are located at times $t_n^{(S)}$ and $t_m^{(S)}$, respectively, exists an oscillating interference contribution $\mathcal{W}_{B,nm}^{(I)}(t,\omega)$ at time \cite{hlawatsch_interference_1997}

\begin{equation}
    t_{nm}^{(I)} = \frac{t_n^{(S)}+t_m^{(S)}}{2},
\end{equation}

which for $|n-m|=q\Delta t$, $q$ being a positive even number, overlaps with the signal term of another pulse in the Wigner space.\\
The Wigner interference pattern shows a discrete symmetry along the frequency axis with period $2\pi/\Delta t$. For each period along the frequency axis, interference in the time-frequency Wigner space leads either to a strong spectral peak or to very weak spectral secondary maxima in between, explaining the burst-typical peak structure in the spectrum. For the time-domain marginal, there are no secondary maxima beside the burst pulses. We note that the Wigner interference contributions $\mathcal{W}_{B,nm}^{(I)}(t,\omega)$ do have a physical significance and are not a mere mathematical artefact. Their total energy can be shown to be zero, because the individual pulse Wigner distributions are time-frequency disjoint (Moyal's formula) \cite{claasen_wigner_1980}. However, they are responsible for the spectral interference structure, which can be measured with any spectrometer with sufficient resolution. Another symmetry property of the Wigner distribution is given by the vertical line at $t=0$, across which it has an even symmetry $\mathcal{W}_B(t,\omega)=\mathcal{W}_B(-t,\omega)$. This symmetry is given by the fact, that we calculated with pulses that are equal in their energy. However, it is also preserved for a nonzero phase slip $\phi_s\neq0$.

\subsection{The low-$N$ regime: Few Strongly Chirped Pulses}
\label{sec:numerical-few}
The chirp parameter $C$ is very high. The chirped pulse duration $\tau_{FWHM}$ is much larger than the pulse spacing $\Delta t$ and also even larger than the whole compressed burst duration $(N-1)\Delta t$

\begin{subequations}
\begin{equation}
    C \gg 1,
\end{equation}
\begin{equation}
    \tau_{FWHM} \gg (N-1)\Delta t > \Delta t
\end{equation}
\end{subequations}

As it is generally known for a single pulse \cite{trebino_frequency-resolved_2000}, the Wigner distribution $\mathcal{W}_{B,chirped}(t,\omega)$ of a chirped burst in this regime (Fig. \ref{fig:wigner-all}, middle) can be seen to be a tilted version of the Wigner distribution of the compressed pulses $\mathcal{W}_{B,compr}$ (Fig. \ref{fig:wigner-all}, left):

\begin{align}
    \mathcal{W}_{B,chirped} (t,\omega) &=
    \mathcal{W}_{B,compr}\left(t-\frac{4\ln{(2)}C}{\omega_{FWHM}^2}\cdot\omega,\omega\right)\nonumber\\ &= 
    \mathcal{W}_{B,compr}\left(t,\omega+\frac{4\ln{(2)}C}{\tau_{FWHM}^2}\cdot t\right)
\end{align}

Giving this tilt in combination with the even symmetry of the compressed-pulse distribution, we note the presence of a diagonal symmetry line in this case, which is in agreement with the mapping of the spectrum into the time domain. Therefore, the burst intensity $I(t)$ is, up to a chirp-dependent factor $a(C)$ in the argument, well represented by the burst spectrum $S(\omega)$

\begin{equation}
    I(t) \propto S\left(a(C)\cdot t\right).
\end{equation}

This is confirmed by our numerical calculations, as can be seen by comparing the Wigner marginal distributions of Fig. \ref{fig:wigner-all}, middle.

\subsection{The high-$N$ regime: Many Strongly Chirped Pulses}
\label{sec:numerical-many}
The chirp parameter $C$ is very high. The chirped pulse duration $\tau_{FWHM}$ and, due to the large number of pulses $N$, the compressed burst duration $(N-1)\Delta t$ are both much larger than the pulse spacing $\Delta t$ 

\begin{subequations}
\begin{equation}
    C \gg 1,
\end{equation}
\begin{equation}
    \tau_{FWHM} \gtrapprox (N-1) \Delta t \gg \Delta t
\end{equation}
\end{subequations}

In this regime, the diagonal symmetry of the chirped low-pulse Wigner distribution is broken due to many chirped-pulse replicas along the horizontal/time axis. When presenting the data while covering a large coordinate range (Fig. \ref{fig:wigner-all}, right), the Wigner distribution can be seen to consist of horizontally spreaded contributions. When taking a closer look at the individual contributions, we see that each contribution consists primarily of diagonal, closely spaced lines (Fig. \ref{fig:wigner-zoom}, right). Wigner signal and interference contributions are hardly distinguishable at this point. Performing the horizontal sum over time to acquire the spectrum, the contributions can be either attributed to signal and constructive interference terms (only positive lines, lower contribution in Fig. \ref{fig:wigner-zoom}, right), retaining the strong spectral peaks, or, to signal and destructive interference terms (positive and negative lines, upper contribution in Fig. \ref{fig:wigner-zoom}, right), suppressing the spectral density in between the peaks. The intensity distribution in time is smoothed out by interference in the time-frequency Wigner space and thus the CPA technique is a useful way to amplify ultrashort-pulse bursts safely without amplifier damage.

\subsubsection{Energy scalability of the high-$N$ regime}
\begin{figure}
\includegraphics[width=1.0\linewidth]{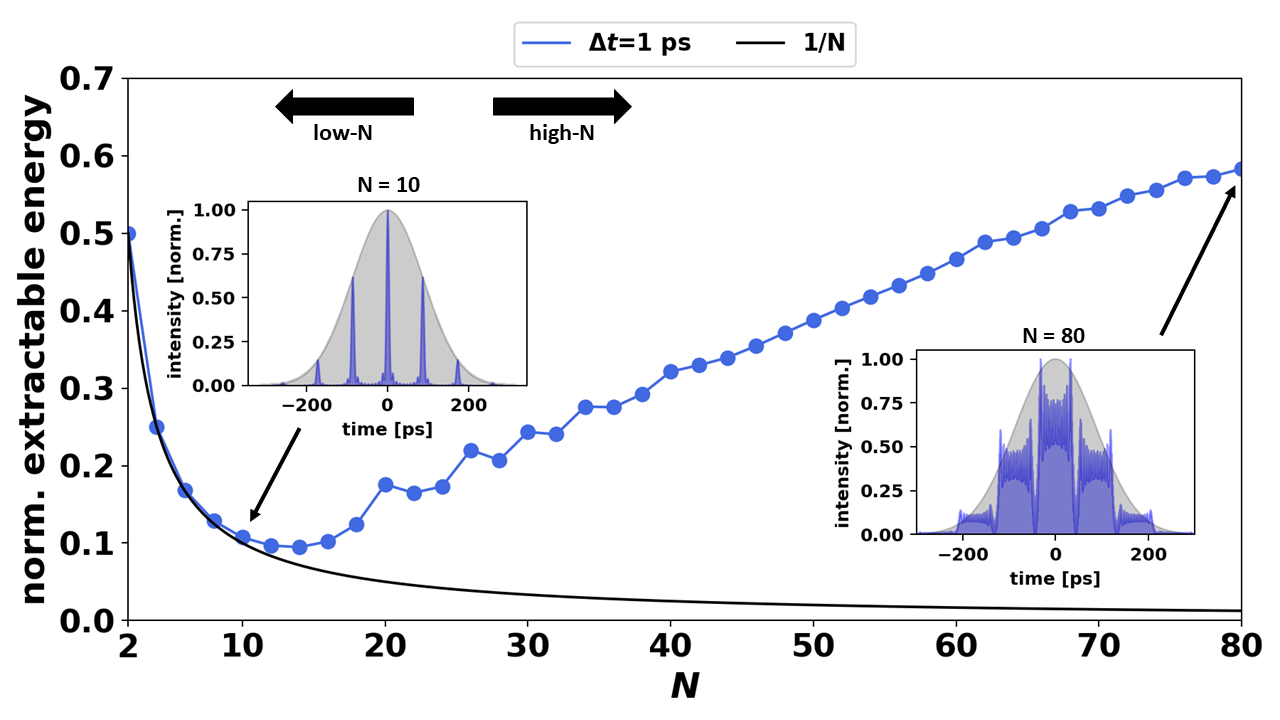}
\caption{\textbf{Normalized extractable energy from an amplifier} at 1 ps interpulse spacing, corresponding to 1 THz burst rate, versus pulse number $N$ (blue). The black solid line shows the $1/N$ scaling behaviour that would apply in the absence of the self-smoothing effect. The inlays show the temporal intensity profile (blue) of chirped bursts at $N=10$ (upper left) and at $N=80$ (lower right) and that of a single chirped pulse (grey), normalized to the intensity threshold $I_{THR}$. Pulse parameters: $C=1.66\cdot10^7, \tau_{FWHM}=200$ ps}
\label{fig:gammas_low}
\end{figure}

Interpulse interferences between strongly chirped pulses lead in the low-$N$ regime to an overshoot of the chirped burst temporal intensity profile. The high-$N$ regime allows for a self-smoothing effect. In this section, we further investigate this phenomenon numerically, with the focus on how much energy can be extracted from an amplifier per amplification cycle, under equal conditions, compared to single-pulse operation when the only limitation is a peak-intensity-induced optical damage threshold $I_{THR}$. We compare the reachable burst energy $\epsilon_B$ with that of a single pulse $\epsilon_P$ at a given intensity threshold $I_{THR}$:

\begin{equation}
    \frac{\epsilon_B}{\epsilon_P} = \frac{\int I_B(t) dt}{\int I_P(t) dt},
\end{equation}

which gives the burst-extractable amplifier energy normalized to single-pulse operation. $I_B(t)$, $I_P(t)$ are the chirped intensity profiles of a burst and a single pulse, respectively, and $\max{\{I_B(t)\}}=\max{\{I_P(t)\}}=I_{THR}$.\\
The results of our numerical investigation can be seen in Fig. \ref{fig:gammas_low} where we show, depending on the number of pulses $N$, the normalized extractable energy of bursts with pulses chirped to 200 ps and an interpulse spacing of 1 ps (corresponding to a 1 THz burst rate). For less than about 10 pulses we see a $1/N$ decrease. This refers to an $N$-times increase of peak intensity of the chirped temporal intensity profile. We note, that this is the same behaviour as from the spectral peak intensity, as can be calculated from Eq. \ref{eq:Eb_omega}. This indicates the discussed frequency-to-time mapping of the spectral peak structure and the absence of the temporal self-smoothing effect. At $N \gtrapprox 10$, we observe a continuous increase in the normalized extractable energy, which is given by the onset of the self-smoothing of the temporal intensity profile of the chirped burst. 

\section{Experimental Setup}
\label{sec:experimental}
The motivation is to directly measure the increasing effect of temporal intensity smoothing in THz-rate bursts of chirped pulses, arising when raising the number of pulses to $N > 10$ pulses. For this we generate and characterize $\mu$J bursts with 1.8 ps spaced pulses (corresponding to a 0.56 THz burst rate) at various pulse numbers. The laser system and an overview of the experimental setup is given in Sec. \ref{sec:experimental-system}. We report on a self-referenced method to stabilize the pulse-to-pulse phase-slip $\phi_s$ in Sec. \ref{sec:experimental-stab}. Spectrogram measurements of the chirped burst by cross-correlation with the synchronous, compressed reference pulse are shown in Sec. \ref{sec:experimental-xcorr}. Finally, we show compressibility of the burst waveform by autocorrelation and techniques for optimization of the burst generation process in Sec. \ref{sec:experimental-AC}.

\subsection{Ultrashort-Pulse Burst Laser System}
\label{sec:experimental-system}
\begin{figure*}[htbp]
\centering
{\includegraphics[width=\linewidth]{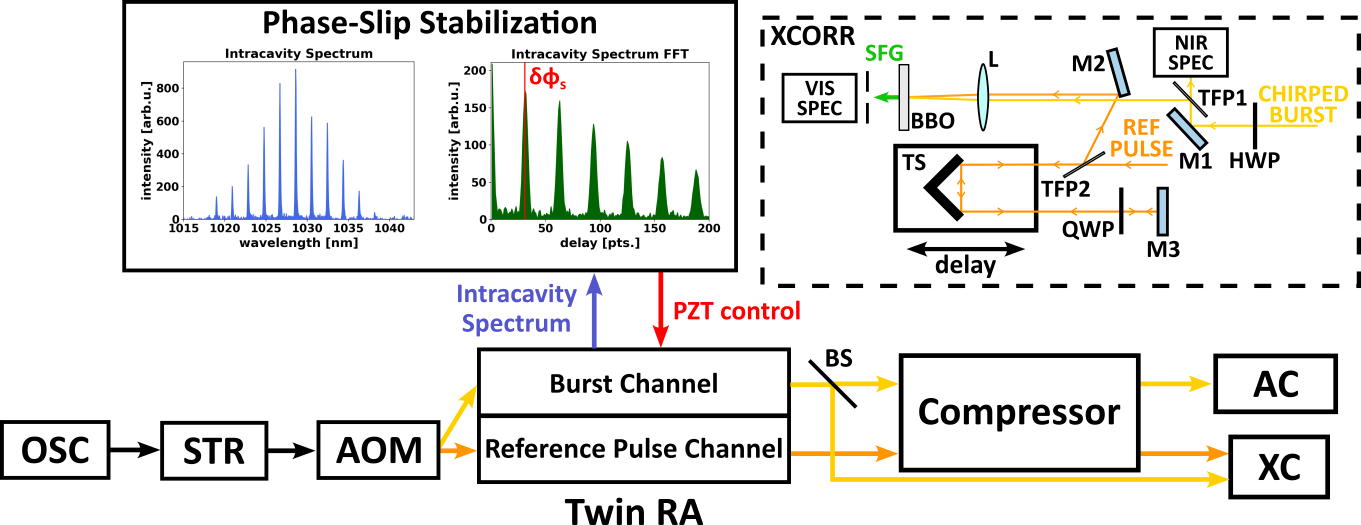}}
\caption{\textbf{Overview of the experimental setup} with a Twin Regenerative Amplifier providing two channels: The burst channel (yellow) and the reference pulse channel (orange). The upper left inlay shows the intracavity spectrum of 40 pulses (blue) and its Fourier Transform (green) including the reference phase (red). The upper right inlay shows the cross-correlation setup in detail. \textbf{OSC:} Oscillator. \textbf{STR:} Stretcher. \textbf{AOM:} Acousto-Optical Modulator. \textbf{RA:} Regenerative Amplifier. \textbf{AC:} Auto-Correlation setup. \textbf{XC:} Cross-Correlation setup. \textbf{TFP1/2:} Thin-Film Polarizer. \textbf{TS}: Translation Stage. \textbf{QWP:} Quarter Waveplate. \textbf{HWP:} Half Waveplate. \textbf{M1/2}: Mirror. \textbf{L:} Lens. \textbf{BBO:} Beta-Barium Borate Crystal. \textbf{VIS SPEC: } Visual Spectrometer. \textbf{NIR SPEC:} Near-Infrared Spectrometer.}
\label{fig:exp}
\end{figure*}
A schematic of the experimental system can be seen in Fig. \ref{fig:exp}. A MHz-repetition-rate mode-locked oscillator (\textbf{OSC}, 1030 nm Yb:KGW, 76 MHz repetition rate, 80 fs pulse duration) generates nanojoule pulses. The oscillator pulses are stretched to around 300 ps by a double-pass grating stretcher (\textbf{STR}). An acousto-optical modulator (\textbf{AOM}) works as pulse picker by diffracting the burst seed pulses. Amplification takes place at a repetition rate of 1 kHz in a CW-pumped twin regenerative amplifier (\textbf{Twin RA}, Yb:CaF$_2$) with two cavities: in one we accumulate the AOM-diffracted pulses to an ultrashort-pulse burst and amplify it up to $\mu$J burst energies, in the other we amplify a single non-diffracted pulse to a few-$\mu$J level as reference for cross-correlation measurements. The prior requires that the round-trip time of the burst cavity is comparable to the oscillator round-trip time, such that their absolute difference gives the intraburst pulse spacing (Vernier effect). By application of an intermediate voltage to the RA Pockels Cell (PC) during burst pulse accumulation, we are able to set round-trip losses and round-trip gain equally, in order to acquire a scalable number of burst pulses with the same energy. In contrast to efforts on Vernier burst generation and amplification in recent times \cite{stummer_programmable_2020}, CPA of ultrashort-pulse bursts does not require any phase-modulation techniques (phase scrambling) due to the smoothing of the temporal intensity peaks in the high-$N$ regime (Fig. \ref{fig:wigner-all} right, Fig. \ref{fig:gammas_low}). Both cavities are seeded by the same oscillator, thus, the burst and the reference pulse are synchronized to each other. The \textbf{Compressor}, containing a single large 130x20 mm$^2$ transmission diffraction grating, compresses both the burst and the reference pulse to 250 fs with beams being spatially separated inside the compressor. Spectra of amplified bursts are measured with a high-resolution Near-Infrared (NIR) spectrometer (\textbf{NIR SPEC}, Avantes Avaspec-ULS4096CL-EVO).

\subsection{Phase-Slip Stabilization}
\label{sec:experimental-stab}
When forming a burst of ultrashort pulses by using the Vernier effect between an oscillator with round-trip time $\tau_{OSC}$ and an RA with round-trip time $\tau_{RA}$, a constant pulse-to-pulse phase slip $\phi_s$ is imposed on the burst pulses, according to

\begin{equation}
    \phi_s = \omega_0 (\tau_{RA} - \tau_{OSC}) = \omega_0 \Delta\tau,
\end{equation}

where $\Delta \tau$ is the round-trip time detuning between RA and oscillator, whose absolute value is equal to the interpulse spacing $\Delta t$. In a first approximation, we assume the phase slip to be constant over the full pulse bandwidth.\\
To generate ultrashort-pulse bursts with stable spectra, drifts in the phase slip induced by drifts in the round-trip time detuning need to be considered. In this work, we apply a slow feedback loop for stabilization of the phase slip without the use of any additional reference to the system. For the pulse-to-pulse phase slip stabilization, we measure the intracavity spectrum of the burst channel (Ocean Optics HR4000). The complex Fast Fourier Transform of the intracavity spectrum shows a modulation peak at a position corresponding to the burst rate (red line in the upper left inlay of Fig. \ref{fig:exp}). Drifts of the phase at this point are equal to drifts of the phase slip. To compensate for any phase variations, we apply a PI control algorithm and change the cavity length accordingly by moving one of the end mirrors by a piezoelectric transducer.

\subsection{Spectrogram Measurements of the Chirped Burst}
\label{sec:experimental-xcorr}
We cross-correlate (\textbf{XC}) the chirped amplified burst with the compressed reference pulse quasi-collinearly (2\degree crossing angle) in a type I Beta-Barium Borate (BBO) crystal and measure the sum-frequency generated (SFG) spectrogram (\textbf{VIS SPEC}, Ocean Optics HR4000CG-UV-NIR), i.e. the spectrum of the SFG signal for each burst-reference time delay point (See Fig. \ref{fig:exp}, upper right inlay). The SFG spectrogram is given as \cite{hlawatsch_interference_1997}

\begin{equation}
    \mathcal{S}_E^{(SFG)}(t,\omega) = \left| \int E_B(t')h(t-t')\exp{(i\omega t')} dt' \right|^2
\end{equation}

with $h(t)$ being a window function, which is given by the compressed reference pulse with 250 fs duration $\tau_{FWHM,ref}$. The spread of the window function $h(t)$ is much smaller ($\tau_{FWHM,ref}$ = 250 fs) than the temporal spacing in all pairs of interfering pulses ($\geq \Delta t$ = 1.8 ps), therefore, interference terms that arise in the burst Wigner distribution are strongly attenuated in the spectrogram. Summation over the wavelength axis gives the time-dependent intensity of the chirped burst waveform. Due to the large duration of the 300 ps chirped pulses, we use a long-range, high-precision translation stage in a double-pass configuration by a combination of thin-film polarizer, quarter waveplate and back-reflecting mirror, to allow for an 800 ps temporal range with 1 ps step size.

\subsection{Compressed Burst SHG Auto-Correlations}
\label{sec:experimental-AC}
We perform Second-Harmonic Generation (SHG) autocorrelation measurements (\textbf{AC}) of the compressed ultrashort-pulse burst in a nonlinear type I BBO crystal. This way, we show compressibility of the burst pulses, without interpulse crosstalk  that distorts the phases of individual amplified pulses. We also use this measurement to set the proper settings for the burst generation process: We find the optimal PC intermediate voltage primarily by two measures

\begin{itemize}
    \item When looking at the burst spectrum, we maximize the ratio of spectral peak height and the spectral background in between the maxima, while keeping the energy constant.
    \item The envelope of the AC should, at its best, be triangular.
\end{itemize}

This way, it is made sure that for a given gain a number of $N$ amplified pulses with equal energies are generated.

\section{Experimental results}
\label{sec:res}
We first validate the cross-correlation method by comparing the result for a single chirped pulse with its spectrum (Sec. \ref{sec:res-xcorr_sp}). Then, we perform measurements for ultrashort-pulse bursts with 20 pulses up to 40 pulses (Sec. \ref{sec:res-xcorr_burst}). Finally, we show performance data of the phase slip stabilization (Sec. \ref{sec:res-stab}) and the autocorrelation of the compressed burst (Sec. \ref{sec:res-AC}).

\subsection{Cross-Correlation Validation with a Chirped Single Pulse}
\label{sec:res-xcorr_sp}
\begin{figure}[htbp]
\centering
{\includegraphics[width=0.6\linewidth]{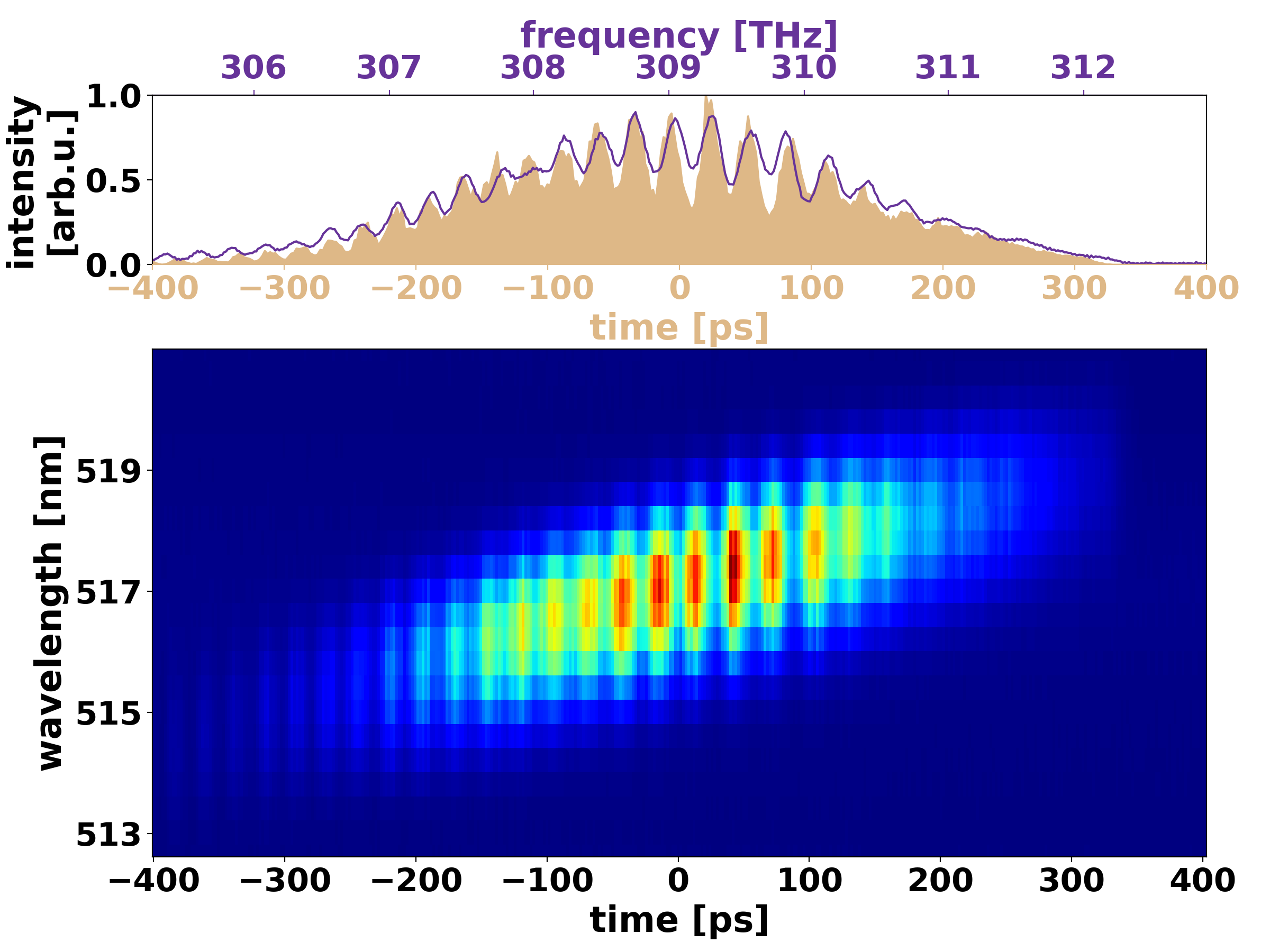}}
\caption{\textbf{Measured spectrogram of a chirped single pulse} by SFG cross-correlation with the compressed reference pulse. Top panel: Intensity over time (brown) acquired by summation over the wavelength axis for each delay point, in comparison with its spectrum (violet).}
\label{fig:sp-xcorr}
\end{figure}

To validate the cross correlation measurement, we compare the measured temporal intensities of a 300 ps chirped single pulse with its spectrum, since it is expected that they are equal to each other for such large chirp parameters ($C > 10^7$). This confirms a good spatial overlap of both channels within the BBO crystal over the whole multiple-100 ps travel range. The result can be seen in Fig. \ref{fig:sp-xcorr}. The spectrum of the original near-infrared (NIR) chirped pulse, is well reproduced by the temporal intensity distribution. Both, the temporal intensity profile and the directly measured spectrum show a periodic modulation structure introduced by the etalon effect in the intracavity air-spaced waveplate. In the spectrogram, the primarily linear chirp is clearly visible, also the chirped pulse duration of 300 ps is confirmed by the time-dependent intensity. 

\begin{figure}
\centering
    \begin{subfigure}[htbp]{0.6\linewidth}
    \centering
    {\includegraphics[width=1.0\linewidth]{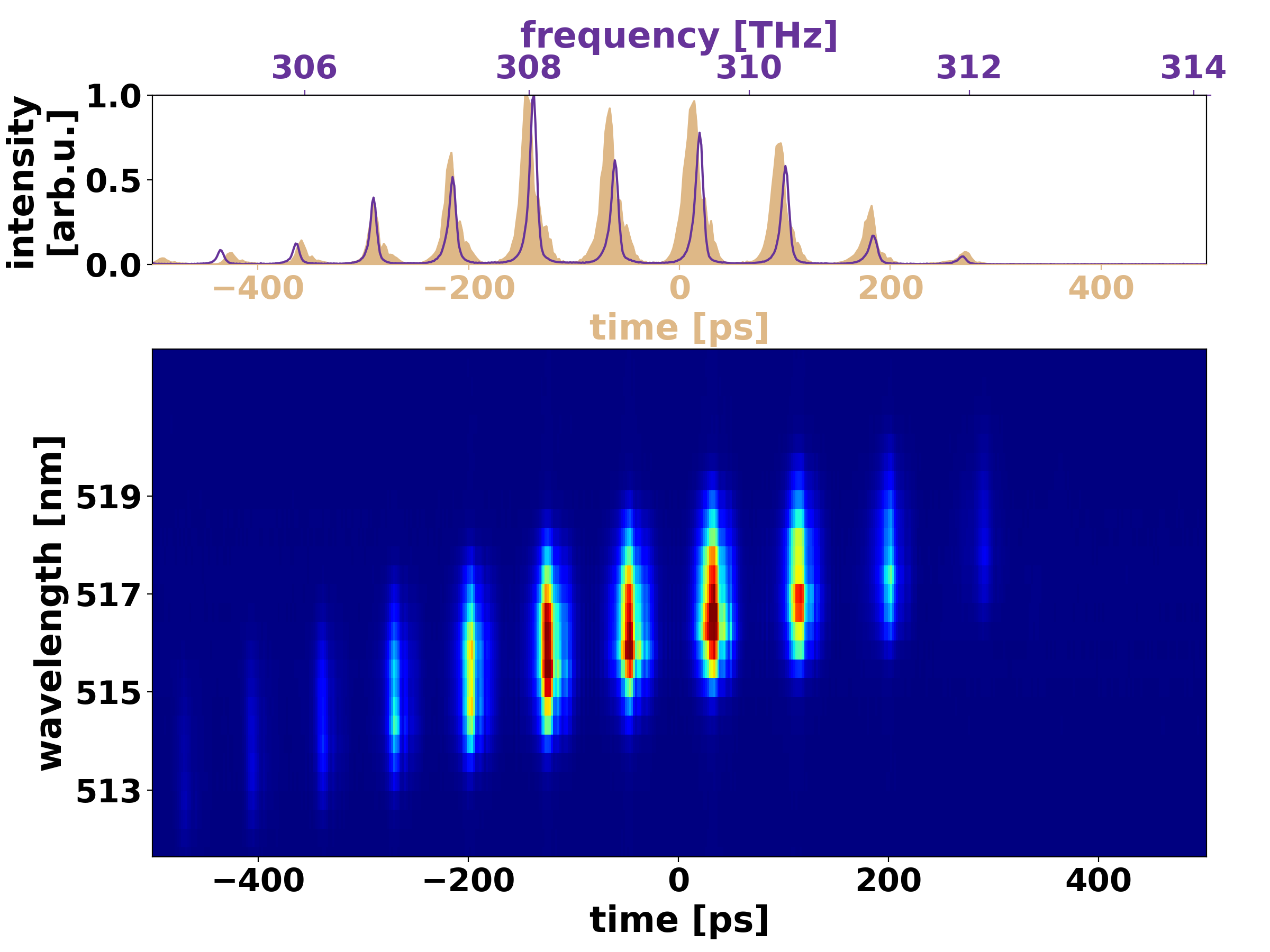}}
    \caption{}
    \label{fig:res-comp-20}
    \end{subfigure}
    
    \begin{subfigure}[htbp]{\linewidth}
    \centering
    {\includegraphics[width=0.6\linewidth]{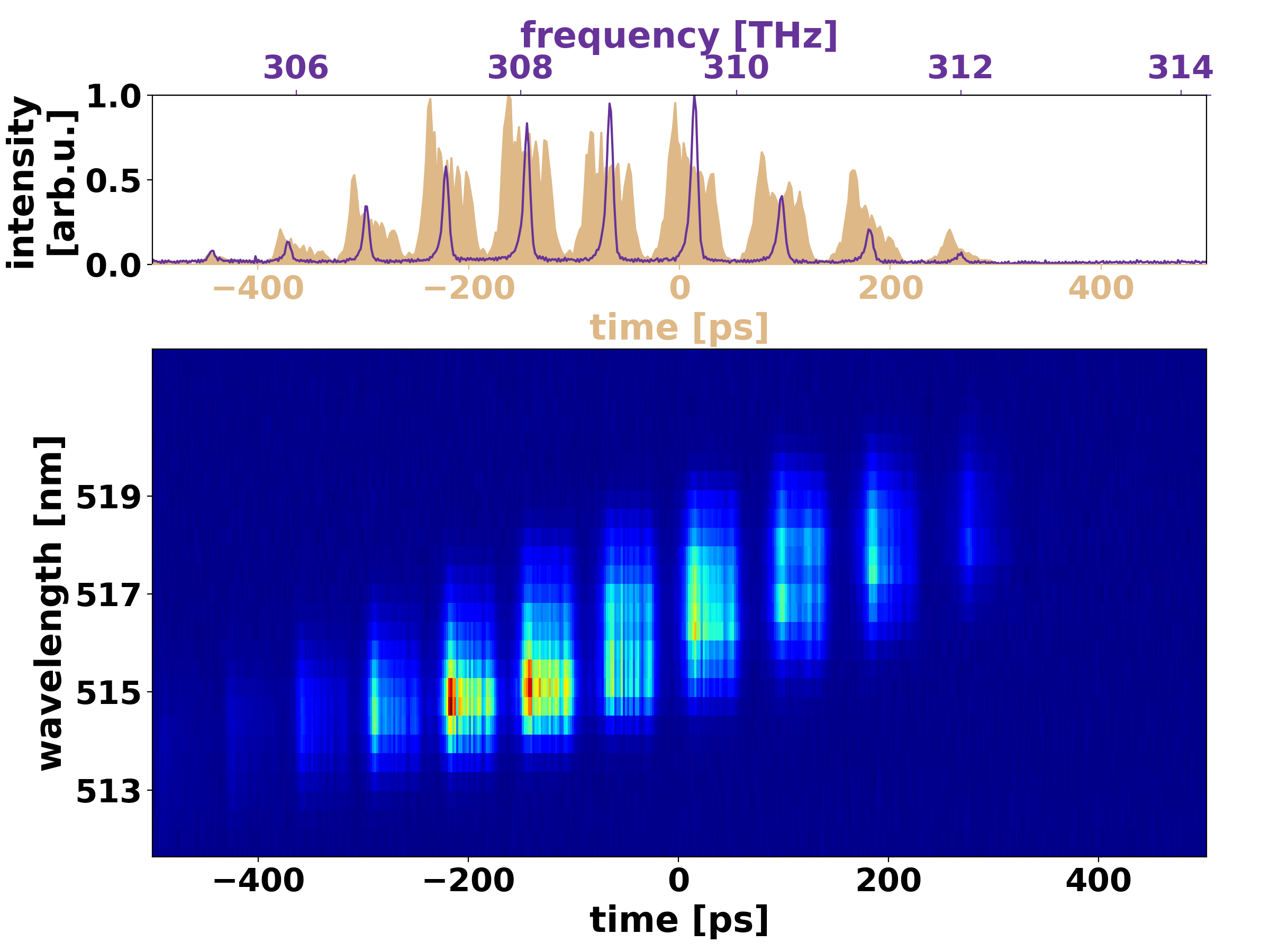}}
    \caption{}
    \label{fig:res-comp-30}
    \end{subfigure}
    
    \begin{subfigure}[htbp]{\linewidth}
    \centering
    {\includegraphics[width=0.6\linewidth]{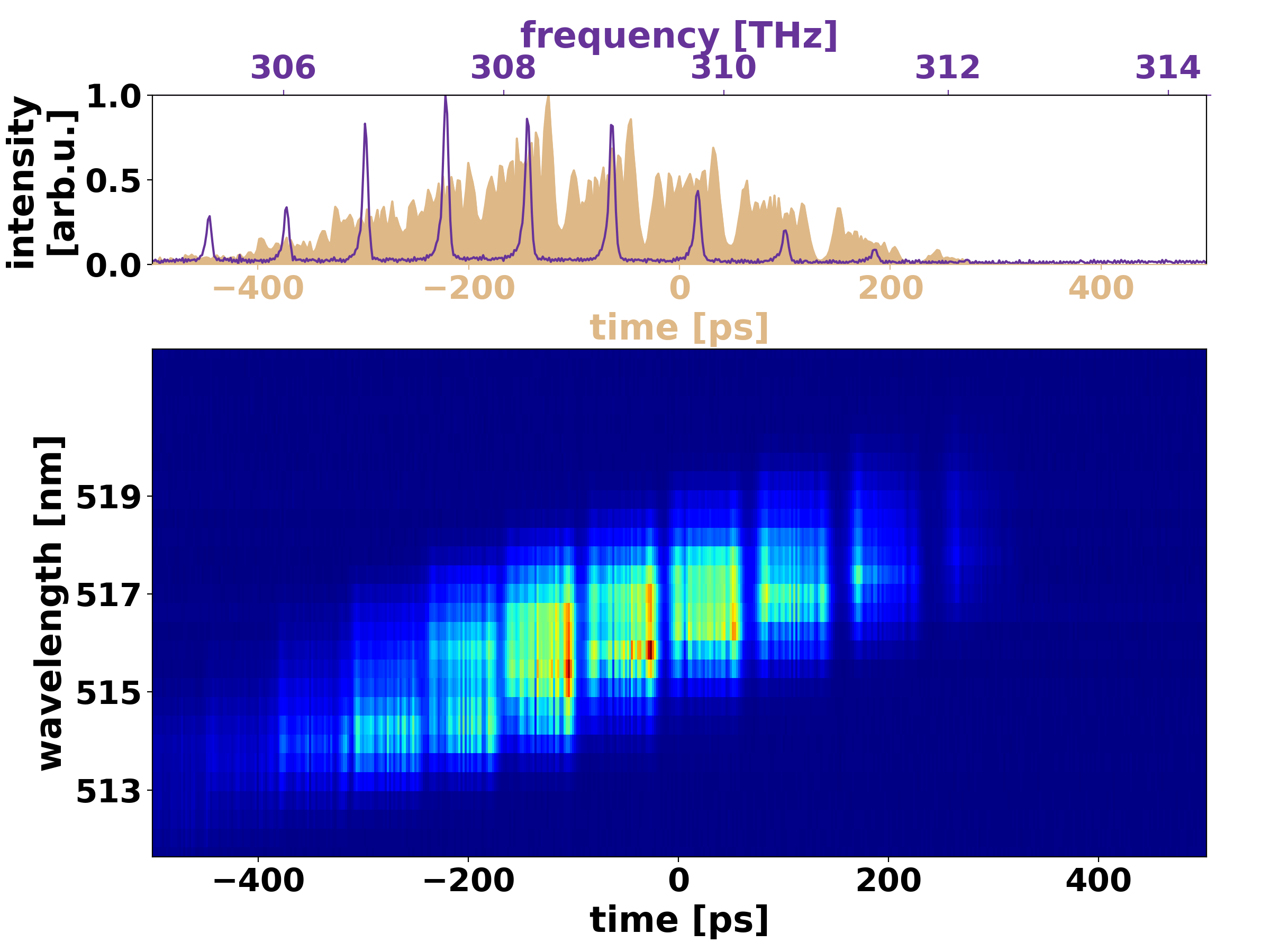}}
    \caption{}
    \label{fig:res-comp-40}
    \end{subfigure}
\caption{\textbf{Measured spectrogram of bursts} consisting of \textbf{(a) 20 (b) 30 (c) 40 chirped pulses} by SFG cross-correlation with the compressed reference pulse. Top panel: Intensity over time (brown) acquired by summation over the wavelength axis, in comparison with its spectrum (violet).}
\label{fig:res-comp}
\end{figure}

\subsection{Cross-Correlation of Chirped Bursts with a Compressed Reference Pulse}
\label{sec:res-xcorr_burst}
\begin{figure}[htbp]
\centering
{\includegraphics[width=0.8\linewidth]{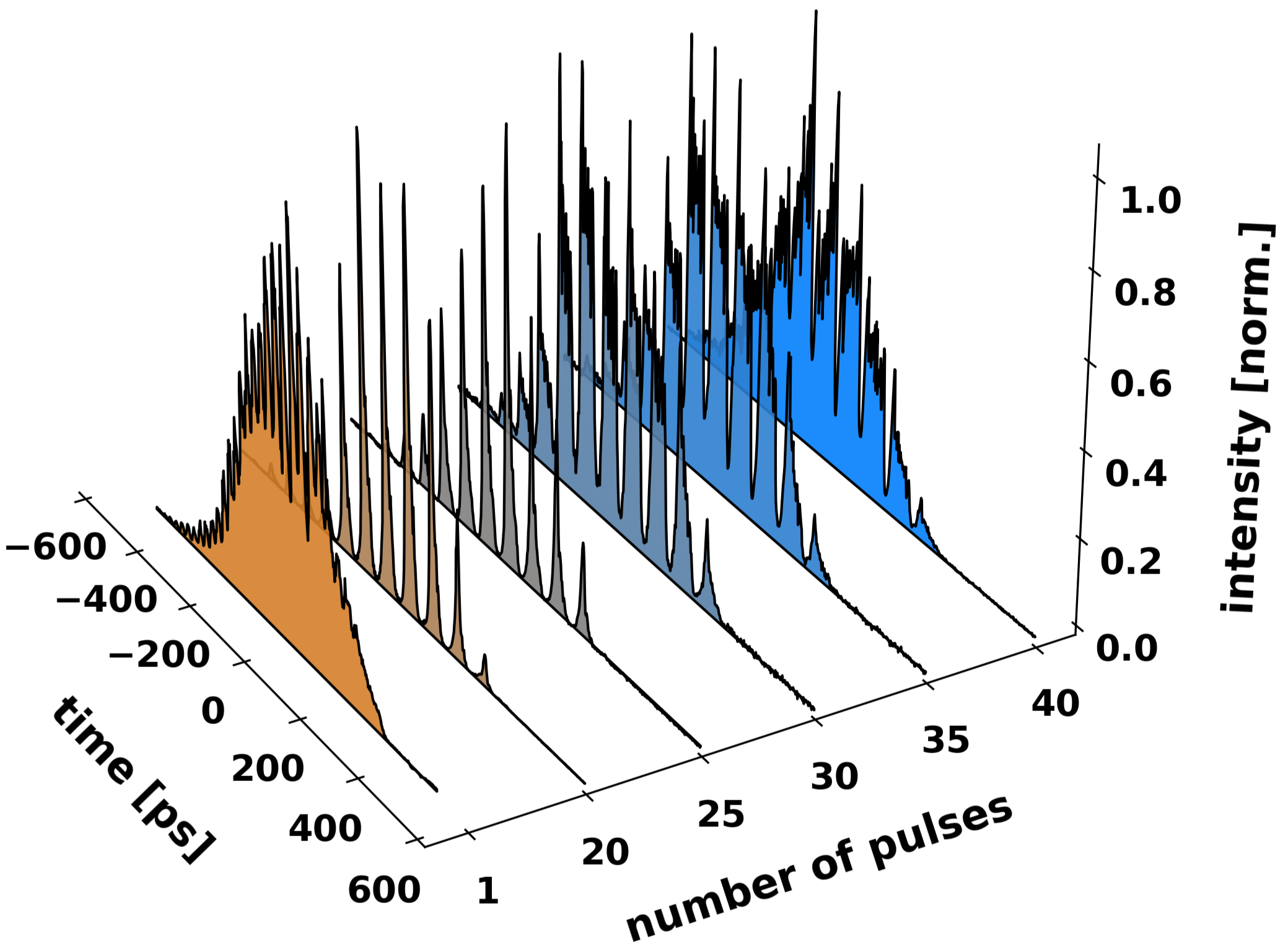}}
\caption{\textbf{Temporal intensity profiles of chirped bursts} acquired from the measured spectrograms by summation over the wavelength axis, shown for a single pulse and for multi-pulse bursts with 20 up to 40 pulses.}
\label{fig:res-comp3d-all}
\end{figure}

We set the gain to a given value in order to get sufficient signal on the SFG spectrometer and in the autocorrelation. For this, we generate bursts with an intraburst pulse spacing of 1.8 ps, corresponding to a 0.56 THz burst rate. We amplify the burst and the reference to about 10 $\mu$J each, and optimized the PC intermediate voltage as describe in Sec. \ref{sec:experimental-AC}. The phase slip stabilization was then turned on. This procedure was done after every time the burst seed pulse number was set by the AOM electronics and the burst accumulation time window in the burst RA channel was adjusted accordingly. We performed measurements from 20 pulses up to 40 pulses in 5-pulse steps.\\
The cross-correlation spectrograms, together with the temporal intensities acquired by summation of the spectrogram data over the wavelength axis are shown in Fig. \ref{fig:res-comp} for 20, 30 and 40 pulses, including a comparison of the acquired temporal intensities with the NIR burst spectra. For 20 pulses (Fig. \ref{fig:res-comp-20}) we see a good agreement of the intensities with the spectra, as it is the case for a single pulses (Sec. \ref{sec:res-xcorr_sp}). This is an indication that for 0.56 THz rate bursts with 20 pulses, which are chirped to 300 ps, we are still in the low-$N$ regime, as discussed in Sec. \ref{sec:numerical-few}. The spectrogram shows, in contrast to the numerically calculated Wigner distributions (Fig. \ref{fig:wigner-all}, middle) only a signal at delay times where peaks in the temporal intensity are visible. In between, no interference structure was recorded. This is in accordance with the formulation of the spectrogram as a smoothed version of the Wigner distribution, where the interference terms are suppressed by a short temporal window. When further increasing the pulse number (Figs. \ref{fig:res-comp-30}, \ref{fig:res-comp-40}), we see a gradual intrinsic smoothing of the signal in time over the whole bandwidth in the spectrogram, which also leads to a smoothing of the time-dependent chirped-burst intensity. We see in Fig. \ref{fig:res-comp3d-all} that the number of 20 pulses seems to be indeed the threshold between the low-$N$ and the high-$N$ regime in our particular case, until at 40 pulses the temporal intensity profile is completely smoothed out. We underline that the high-$N$ regime thus serves for the optimal conditions for CPA.

\subsection{Phase-Slip Stabilization}
\label{sec:res-stab}

\begin{figure}[h]
\centering
{\includegraphics[width=0.8\linewidth]{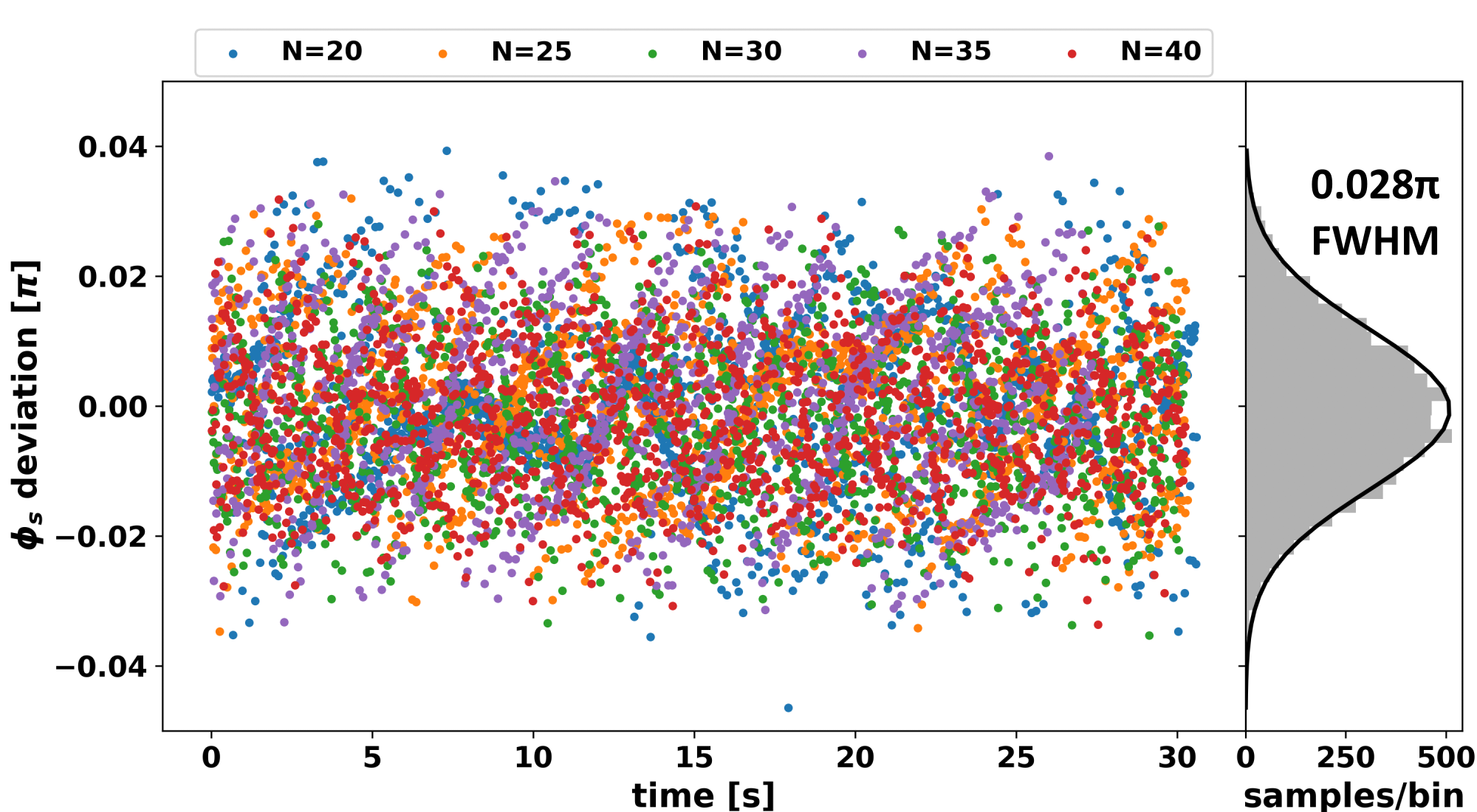}}
\caption{\textbf{Time-dependent deviation of the phase slip $\phi_s$} from an arbitrary target value for 20 up to 40 pulses. On the right side, we show the stochastic distribution by summation of all data points over time (gray) including a Gaussian fit with a 0.028$\pi$ FWHM width (black).}
\label{fig:phasesVtime}
\end{figure}
\begin{figure}[h]
\centering
{\includegraphics[width=0.8\linewidth]{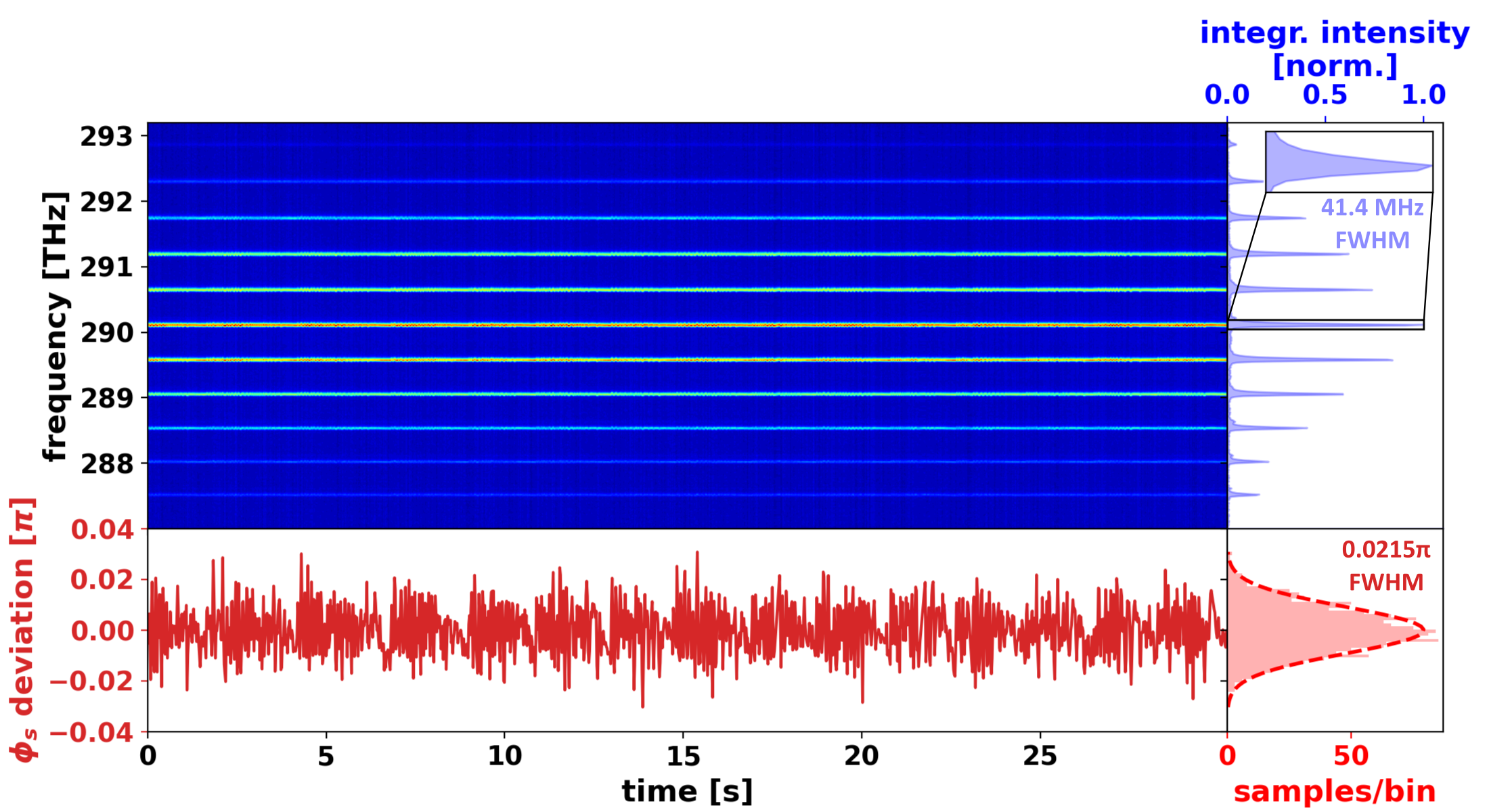}}
\caption{\textbf{Intracavity spectrum from the RA burst channel cavity} over time (top) and the derived time-dependent deviation of the phase slip $\phi_s$ (bottom).}
\label{fig:REFspecVstime}
\end{figure}

\begin{figure*}
\centering
{\includegraphics[width=0.99\linewidth]{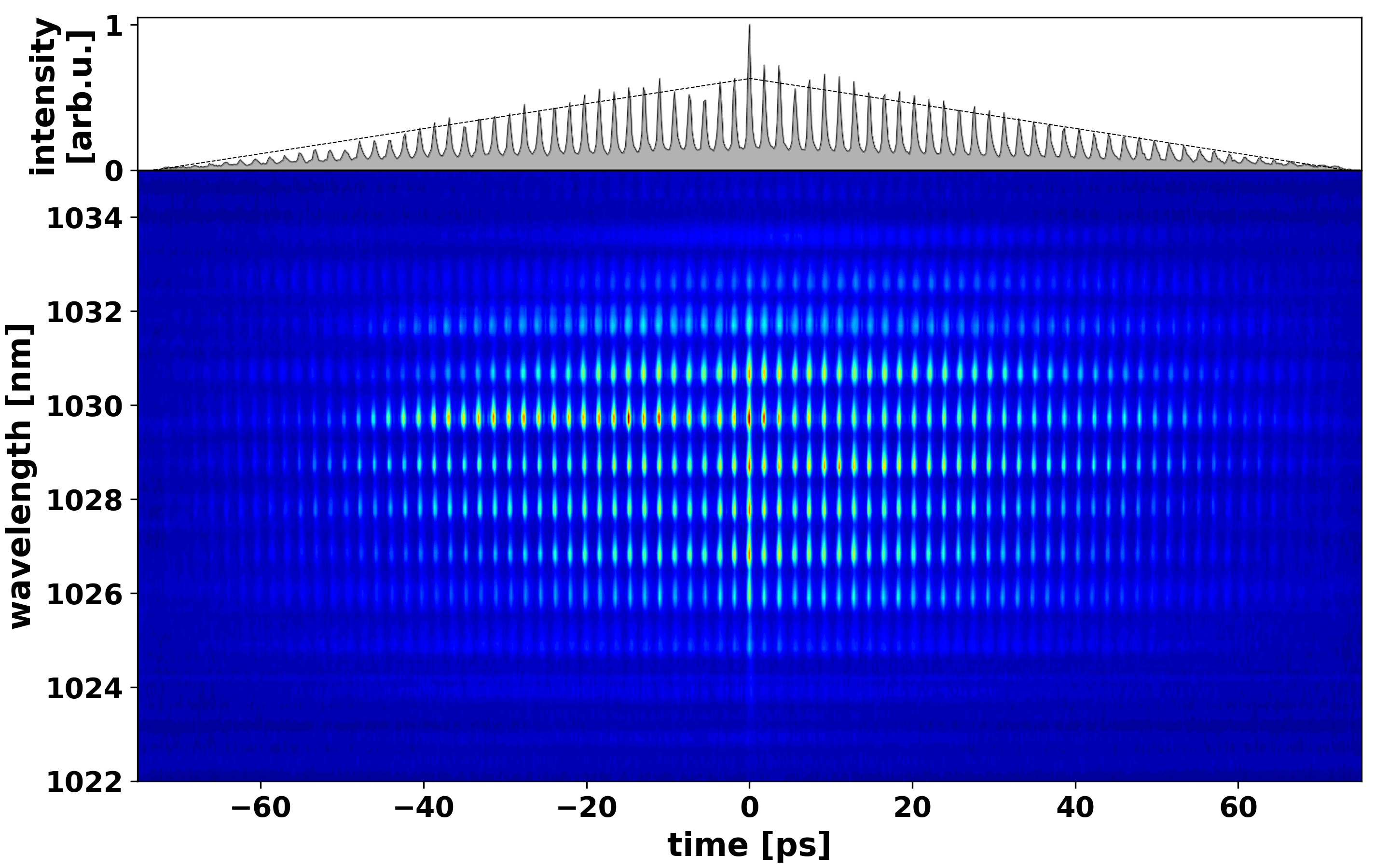}}
\caption{\textbf{Characterization of a 40-pulse 0.56 THz burst at $\mu$J level}: The burst consists of compressed pulses with 250 fs pulse duration. The SHG FROG trace (bottom) and the derived autocorrelation by summation over the wavelength axis (top). The dashed line shows the triangle envelope of the peak maxima (ignoring the zero-delay artifact).}
\label{fig:AC40p}
\end{figure*}

In the following, we present the results for the self-referenced phase-slip stabilization, as it is described in Sec. \ref{sec:experimental}. We recorded the stabilization data in parallel to the acquisition of the cross-correlation data (see Sec. \ref{sec:res-xcorr_burst}).\\
In Fig. \ref{fig:phasesVtime}, we show the measured deviation of the phase slip $\phi_s$ over time. The stochastic distributions of the time-dependent deviations do not depend strongly on the pulse number, which is why we summed up all data points over the time axis. We see that the total phase deviation distribution is Gaussian with an FWHM width of 0.028$\pi$, which corresponds at 1030 nm to an FWHM group delay deviation of only 48 as in the pulse spacing. The phase-slip stabilization performs even well also at high pulse numbers. For 40 pulses, we show the intracavity spectrum over time and its derived phase deviations in Fig. \ref{fig:REFspecVstime}. We use a guided PZT with a 40 $\mu$m travel range (Piezosystem Jena PU 40) that is much larger than the 14.4 nm translation that would correspond to the measured FWHM width of the phase deviation distribution. A main factor for the stochastic width of the phase deviation is given by the limited control loop bandwidth. Our slow control loop was running software-controlled with a mean sample rate of 49 Hz, which could be easily improved by applying a fast hardware-running control loop in the future. Because of the good stability of the phase slip, the peak structure in the burst spectrum can be seen to be very stable. 

\subsection{Autocorrelation of the Compressed Burst}
\label{sec:res-AC}

We show compressibility of the burst pulses by demonstration of SHG autocorrelation results for compressed bursts, which in the case of $N$ pulses is supposed to show $2N-1$ signal peaks. The result for 40 pulses can be seen in Fig. \ref{fig:AC40p}, including the SHG FROG trace from which we derived the autocorrelation. The SHG FROG trace proves further our phase slip control stability by a stable peak structure over time without any noticeable wavelength-detuning drift. The autocorrelation can be seen to be typical for a complex waveform with a broad background component and a coherent artifact \cite{ratner_coherent_2012,raymer_complex_1994}, visible by the overshoot at zero time delay. The peak maxima can be approximately fitted with a triangle-shaped line, indicating equalization of the burst pulse energies. Deviations from the ideal triangle course do not only arise because of the pulses themselves, but also because of the set time delay step of 200 fs of the delay stage that is comparable to the compressed pulse duration of 250 fs, leading to discretization errors. Nonetheless, this delay step value is found to be a good trade-off between step size and a large, almost 160 ps, scan range for our proof-of-concept experiment. We also note, that bursts with any higher pulse number can be easily generated with the described method, however, due to limited scan range and signal-to-noise ratio (SNR) in the AC, higher pulse numbers than 40 were not reasonable for the given demonstration. The theoretical limit of the burst duration is given by the RA burst channel cavity round-trip time, which is approximately 13 ns in our case (corresponding to the 76 MHz oscillator repetition rate). This would correspond to a burst of 13,000 pulses at a 1 THz burst rate. 

\begin{figure}[h]
\includegraphics[width=0.8\linewidth]{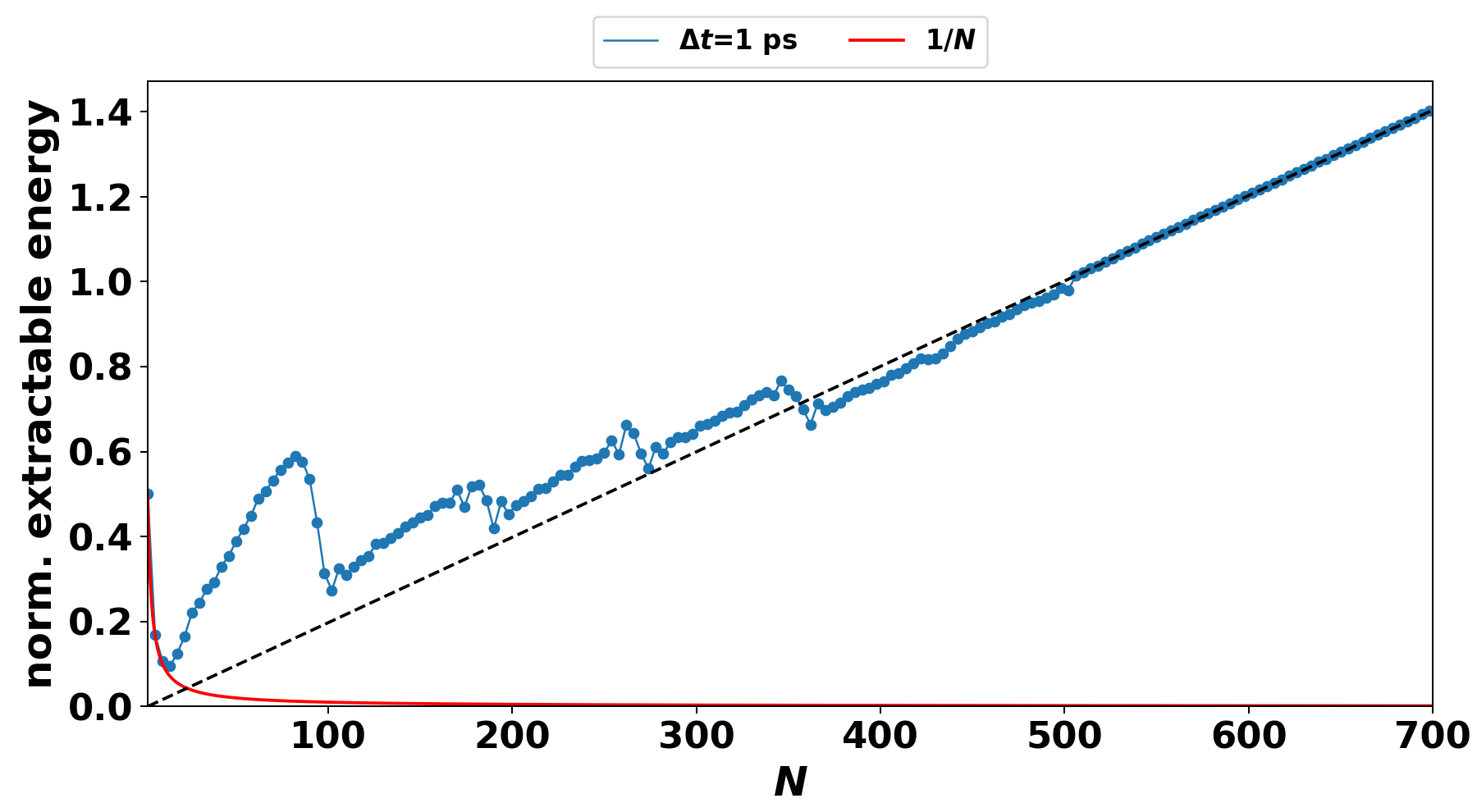}
\caption{\textbf{Normalized extractable amplifier energy for burst-mode operation compared to single-pulse operation} depending on pulse number $N$, at a given intensity damage threshold $I_{THR}$. The black solid line shows the $1/N$ scaling behaviour that would apply in the absence of the self-smoothing effect. The dashed line in black indicates when the maximum interpulse spacing $(N-1)\Delta t$ becomes larger than the duration of an individual chirped pulse $\tau_{FWHM}$. Pulse parameters: $C=1.66\cdot10^7, \tau_{FWHM}=200$ ps.}
\label{fig:gammas}
\end{figure}

\begin{figure}[h]
\includegraphics[width=0.8\linewidth]{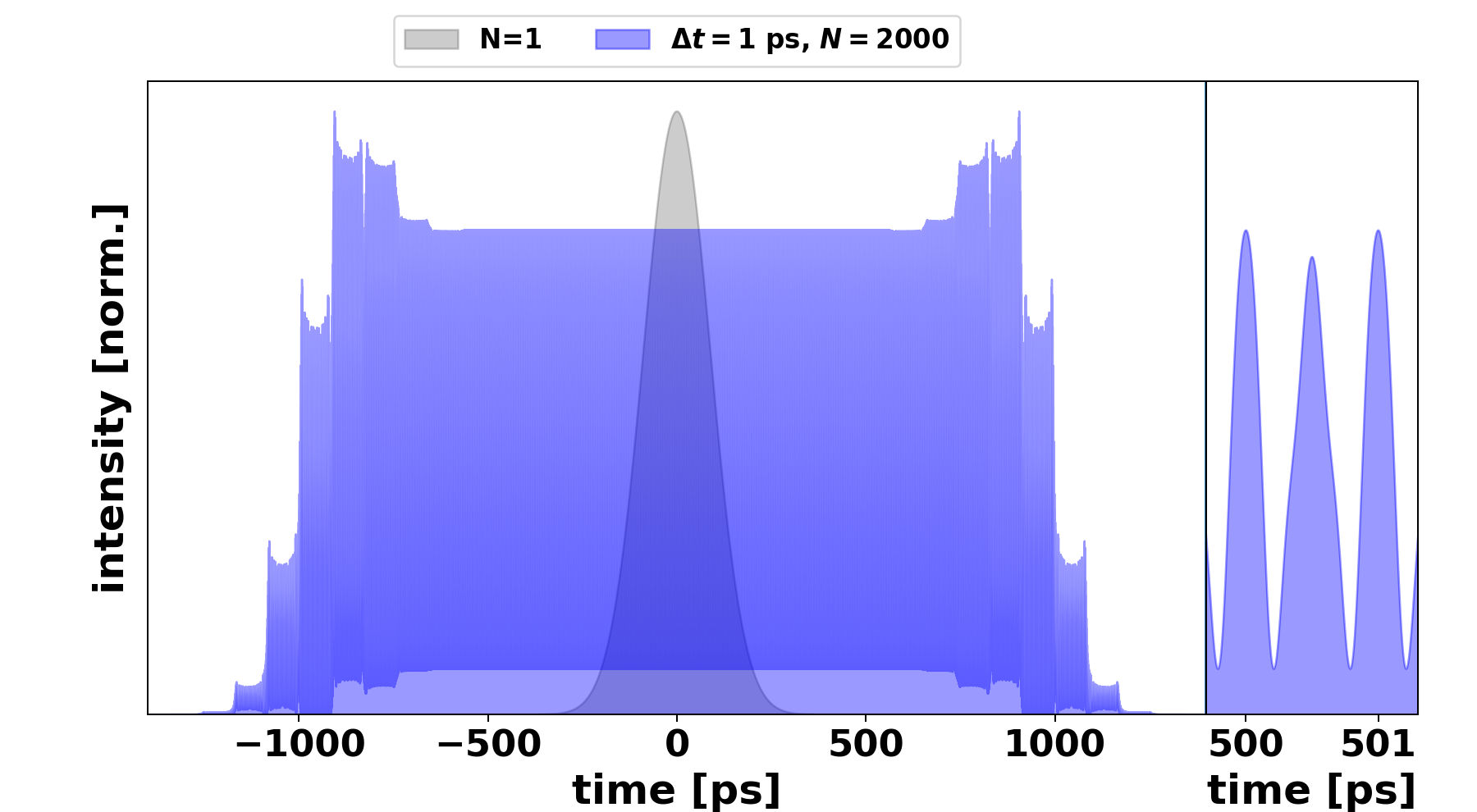}
\caption{\textbf{Normalized temporal intensity profile of the chirped waveform} in the case of a single pulse (grey), $N=2000$ pulses spaced by $\Delta t=1$ ps (light blue). Pulse parameters: $C=1.66\cdot10^7, \tau_{FWHM}=200$ ps.}
\label{fig:temp-profile}
\end{figure}

\section{Outlook on energy scalability for extraordinarily high pulse numbers}
The primary focus of this work is to investigate, both numerically and experimentally, the onset of the high-$N$ regime given by the intrinsic self-smoothing of the chirped temporal intensity profile. In order to give a further outlook on the capabilities given by direct time-domain generation of bursts and the self-smoothing phenomenon at THz burst rates, we further investigate the normalized exctractable energy at pulse numbers $N\rightarrow 1000$ at a 1 THz burst rate.\\
We see a partial periodic revival of the temporal peak structure at every 90 pulses, i.e. at $N=100,N=190,N=280,...$, which is indicated by a decrease in extractable burst energy. The energy decrease due to the partial peak revivals, however, becomes less the higher the number of pulses. In consequence, normalized extractable energy keeps increasing for sufficiently many pulses, until it becomes linearly dependent on the pulse number $N$.\\
For sufficiently large $N$, the largest interpulse spacing $(N-1)\Delta t$ (which is the spacing between the first and the last burst pulse) becomes larger than the duration of the individual chirped pulses $\tau_{FWHM}$ (200 ps in our simulation). In this case, the burst-extractable energy is higher than the extractable energy with a single pulse under the same conditions. This can be well understood, when considering the temporal intensity profile of a chirped burst with $N \rightarrow 1000$, or more, as it is shown in Fig. \ref{fig:temp-profile} with $N=2000$ pulses. In this case, we have a 10-times higher maximum interpulse spacing than the chirped pulse duration. The intensity profile of a strongly chirped THz-rate burst almost completely fills out the intensity-time area, resembling a waveform with a rectangular shape. It also exceeds the temporal range of the single stretched pulse and thus allows for higher energies at a given chirp rate $C$. When zooming into the waveform (right subplot of Fig. \ref{fig:temp-profile}), a periodic temporal peak structure can be observed, with a period equal to the interpulse spacing $\Delta t$. The reason for the normalized extractable energy increase is similar to that of Divided Pulse Amplification (DPA) \cite{zhou_divided-pulse_2007}. In DPA, the total amplified energy is distributed over time over multiple compressed pulses, while peak intensity is kept below a given damage threshold. We note, that the regime $N\rightarrow 1000$, or higher, thus includes advantages of DPA and CPA. 

\section{Conclusion}
We have investigated numerically and experimentally the regime of ultrashort-pulse THz-rate bursts at high ($N\gg10$) pulse numbers, with a focus on the transition from few to many pulses where we observed a gradual intrinsic smoothing of the temporal intensity profile of a chirped burst. Direct self-stabilized burst generation allows for THz burst rates with stable, MHz-wide spectral peaks given by bursts with high pulse numbers. This allows for generation of a controlled stable peak structure that is useful for many nonlinear spectroscopic applications. The pulse-to-pulse phase slip $\Delta \phi_s$ can be stabilized without any external reference to a high degree with an FWHM phase deviation of down to about 0.02$\pi$, as was shown in this work. Phase stability can be characterized by the spectral peak linewidth for which high-resolution spectrometers are available. In the high-$N$ regime, the presence of temporal intensity spikes in the chirped burst waveform are avoided due to the self-smoothing effect. For sufficiently high pulse numbers $N\rightarrow1000$, the largest interpulse burst spacings exceed the chirped pulse duration at THz burst rates. This leads to a combination of CPA and DPA methodologies and to burst-extractable energies from the amplifer that are higher than the extractable energy with a single pulse, under the same conditions. We underline that conventional master-oscillator regenerative-amplifier systems may easily be able to apply this technique with only minor modifications, that are installation of a 3-level PC voltage driver and the adaption of the oscillator or amplifier round-trip time for acquiring the desired intraburst pulse spacing $\Delta t$. Future efforts will include suitable burst characterization techniques that allow for precise determination of the number of pulses and their energies over nanosecond burst durations when scaling the pulse number higher than experimentally demonstrated. Further, generation of bursts with thousands of pulses may require realization of a dispersion-free RA cavity. Overall, these findings strongly underline the significance of the high-$N$ regime for future developments in high-energy ultrashort-pulse burst technology.

\section*{Supplementary Material}
See the supplementary material for detailed derivations of the analytical formulations.

\section*{Conflict of Interest}
The authors have no conflicts to disclose.

\section*{Funding}
\"Osterreichische Forschungsf\"orderungsgesellschaft (I 5590).


%
%

%


\bibliographystyle{unsrt} 
\bibliography{chirped-burst}

\end{document}